\documentclass[%
reprint,
amsmath,amssymb,
aps,
pra,
floatfix,
]{revtex4-2}

\usepackage{graphicx}
\usepackage{dcolumn}
\usepackage{bm}

\begin{document}
\setlength\textfloatsep{6pt}


\title{Heralded single-photon source based on superpositions of squeezed states}

\author{Hiroo Azuma$^{1}$}
\email{zuma@nii.ac.jp}
\author{William J. Munro$^{2}$}
\author{Kae Nemoto$^{2, 1}$}
\affiliation{$^1$Global Research Center for Quantum Information Science,
National Institute of Informatics, 2-1-2 Hitotsubashi, Chiyoda-ku, Tokyo 101-8430, Japan}
\affiliation{$^2$Okinawa Institute of Science and Technology Graduate University, Onna-son, Okinawa 904-0495, Japan}




\date{\today}

\begin{abstract}
We propose a heralded single-photon source based on injecting a superposition of oppositely squeezed states onto a beam splitter.
Our superposition of squeezed states is composed of only even photon number states
(the number of photons is equal to $2,6,10,...$)
meaning the probability for an emitted single photon given as a heralded single-photon event is higher than what one can achieve from the usual two-mode squeezed state.
This enables one to realize an enhanced
heralded single-photon source.
We discuss how to create this superposition of squeezed states utilizing a single-mode squeezed state and the cross-Kerr nonlinearity.
Our proposed method
significantly improves the probability of emitting the heralded single photon compared to spontaneous parametric down-conversion.
\end{abstract}

\maketitle


\section{\label{section-introduction}Introduction}
Single-photon sources are essential tools in many quantum information-based technologies including quantum key distribution protocols \cite{Bennett1984}
and linear optical quantum computation \cite{Knill2001,Kok2007}.
As such there has been
significant
world-wide effort to propose and demonstrate on-demand single-photon emitters.
Quantum dots \cite{Benson2000,Michler2000,Santori2001,Pelton2002,Kiraz2004},
nitrogen-vacancy centers in diamond \cite{Bernien2012,Albrecht2013,Liebermeister2014},
and negative silicon vacancy centers in diamond \cite{Rogers2014a,Rogers2014b,Benedikter2017} have been proposed as solid state devices for realizing such emitters.
Further cavity QED systems have been considered as a triggered single-photon source \cite{Kuhn1999,Brattke2001,Mucke2013,Azuma2019},
but the current work horse is still spontaneous parametric down-conversion (SPDC) where we trigger off the idler photon
\cite{Fasel2004,Mosley2008,Brida2012,Krapick2013,Ngah2015,Kaneda2015}.
However,
the photon pair emission rate is low
[$4\times 10^{-6}$ per pump photon
using a
periodically poled lithium niobate (PPLN) waveguide \cite{Bock2016}].

Implementation of heralded single-photon sources with the SPDC is state-of-the-art technology and is being studied with great effort currently
\cite{Kiessler2023}.
A review of the heralded single-photon sources was given in \cite{Castelletto2008,Meyer-Scott2020}.
Improvement of the heralded single-photon source with photon-number-resolving detectors was examined
in Refs.~\cite{Davis2022,Stasi2023}.
Suppression of multiphoton events in a cavity-enhanced SPDC via the photon-blockade effect for the heralded single-photon source has been proposed \cite{Tang2021},
and a technique of heralding based on the detection of zero photons was investigated \cite{Nunn2021}.

The essence of the heralded single-photon source is the generation of pairs of
photons and is the
reason we use SPDC.
A more efficient method for generating entangled photons is the injection of squeezed light into a 50-50 beam splitter \cite{Kim2002}.
The resulting two-mode squeezed state source can in principle generate
pairs of photons with a much higher success probability.
Using an imperfect single-photon detector whose efficiency is equal to $0.9$,
the pure two-mode squeezed state with a squeezing parameter $0.5$ gives the photon pair emission rate
$0.151$.
There is, however, still room for improvement. 

In this paper,
we introduce a more efficient approach based on superpositions of squeezed states.
We
investigate how to generate a pair of photons by injecting an odd superposition of squeezed states
onto a 50-50 beam splitter \cite{Marek2006}.
In particular
we show that we can create this superposition
using a single-mode squeezed state source and a cross-Kerr nonlinear medium,
although its heralded success probability is less than $1/2$
\cite{Sheng2008,Ding2017}.
Finally, we determine the second-order intensity correlation functions $g^{(2)}(0)$ for the single-photon emissions
of the superposition of oppositely squeezed states and the two-mode squeezed state.
We use the measure $g^{(2)}(0)$ to evaluate whether or not the physical properties of a given photon source
are close to those of an on-demand single-photon source \cite{Walls1994}.

This article is organized as follows.
In Sec.~\ref{section-first-method}, we begin by discussing the generation of a pair of photons using our superposition of oppositely squeezed states
determining in Secs.~\ref{subsection-generation-entangled-states} and \ref{subsection-numerical-calculations}
the probability and conditional probability of emission of the heralded single-photon source.
Then in Sec.~\ref{subsection-errors},
we consider errors induced by imperfect operations of devices.
Next in Sec.~\ref{section-second-method},
we
compare our proposed heralded source with that
from the pure two-mode squeezed state.
In Sec.~\ref{section-second-order-intensity-correlation-functions},
we determine the second-order intensity correlation functions for our
source using
imperfect single-photon detectors.
Finally in Sec.~\ref{section-discussion} we provide a brief concluding discussion.

\section{\label{section-first-method}Photon pair generation from a superposition of oppositely squeezed states}
\begin{figure}
\begin{center}
\includegraphics[width=0.9\linewidth]{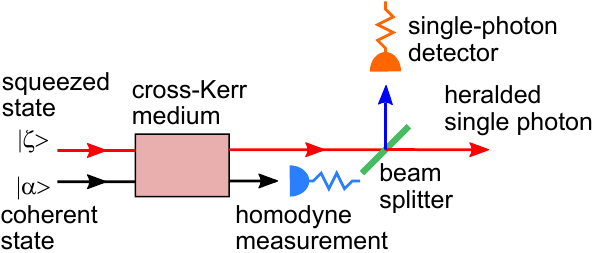}
\end{center}
\caption{Schematic illustration of a heralded single-photon source using a squeezed state and a cross-Kerr medium.}
\label{figure01}
\end{figure}

Let us begin by discussing our scheme
for producing the superposition of the oppositely squeezed states
using a nonlinear cross-Kerr medium.
This is depicted in Fig.~\ref{figure01}.
First, we prepare
a two-mode initial state where the
first and second modes are given by a squeezed state $|r\rangle=\hat{S}(r)|0\rangle$ and a coherent state $|\alpha\rangle$, respectively,
as
\begin{equation}
|r\rangle_{1}
=
\frac{1}{\sqrt{\cosh r}}
\sum_{n=0}^{\infty}
\frac{\sqrt{(2n)!}}{2^{n}n!}
(-\tanh r)^{n}
|2n\rangle_{1},
\end{equation}
\begin{equation}
|\alpha\rangle_{2}
=
\exp(-|\alpha|^{2}/2)
\sum_{n=0}^{\infty}
\frac{\alpha^{n}}{\sqrt{n!}}|n\rangle_{2},
\end{equation}
where $\hat{S}(r)$ is the squeezing operator and $\{|n\rangle:n=0,1,2,...\}$ represent
the photon number states.
Both of these states then interact with a nonlinear cross-Kerr interaction
of the form
$\hat{H}_{12}
=
\hbar\kappa
\hat{a}_{1}^{\dagger}\hat{a}_{1}
\hat{a}_{2}^{\dagger}\hat{a}_{2}$,
where $\hat{a}_{1}$ and $\hat{a}_{2}$ are annihilation operators of modes $1$ and $2$, respectively.
Formally the squeezing operator can be written as $\hat{S}(r)=\exp[(-r/2)(\hat{a}^{\dagger 2}-\hat{a}^{2})]$.
Our initial state $|\Psi_{\mbox{\scriptsize in}}\rangle_{12}=|r\rangle_{1}|\alpha\rangle_{2}$ evolves over a time $\tau$ to
\begin{eqnarray}
|\Psi_{\mbox{\scriptsize out}}\rangle_{12}
&=&
\frac{1}{\sqrt{\cosh r}}
\sum_{n=0}^{\infty}
\frac{\sqrt{(2n)!}}{2^{n}n!}
(-\tanh r)^{n} \nonumber \\
&&
\times
|2n\rangle_{1}|\alpha e^{-i2n\kappa\tau}\rangle_{2}.
\label{time-evolution-cross-Kerr-nonlinear-interaction-0}
\end{eqnarray}
Setting
$2\kappa\tau=\pi$
we have
\begin{eqnarray}
|\Psi_{\mbox{\scriptsize out}}\rangle_{12}
&=&
\frac{1}{2}{\cal N}_{+}(r)^{1/2}
|r;+\rangle_{1}|\alpha\rangle_{2} \nonumber \\
&&
+
\frac{1}{2}{\cal N}_{-}(r)^{1/2}
|r;-\rangle_{1}|-\alpha\rangle_{2},
\label{superposition-zeta-minus-zeta-0}
\end{eqnarray}
where
\begin{equation}
|r;\pm\rangle
=
\frac{|r\rangle\pm|-r\rangle}{\sqrt{{\cal N}_{\pm}(r)}},
\end{equation}
with
\begin{equation}
{\cal N}_{\pm}(r)
=
2
\left(
1\pm
\frac{1}{\cosh r\sqrt{1+\tanh^{2}r}}
\right).
\end{equation}
The states $|r;+\rangle$ and $|r;-\rangle$ are superpositions of the
photon number states
$\{|2n\rangle: n=0,2,4,...\}$ and $\{|2n\rangle: n=1,3,5,...\}$, respectively.

Next, distinguishing $|\alpha\rangle_{2}$ and $|-\alpha\rangle_{2}$
by homodyne measurement,
we
project our system to either
$|r;+\rangle$ or $|r;-\rangle$
with probability ${\cal N}_{\pm}(r)/4$, respectively.
In general, $|\alpha\rangle$ and $|-\alpha\rangle$ are never perfectly orthogonal
but $|\langle\alpha|-\alpha\rangle|^{2}\leq 0.0012$ for $|\alpha|>1.3$.
Here, we assume that $|\alpha|$ is sufficiently large.
With the squeezing parameter $r=0.725$ (corresponding to a squeezing level of $6.3$ dB),
${\cal N}_{+}(r)/4$ and ${\cal N}_{-}(r)/4$ are equal to $0.833$ and $0.167$, respectively.
It is worth mentioning that this probability ${\cal N}_{-}(r)/4$ monotonically increases from $0$
to
$1/2$
as $r$ goes to infinity.

In our scheme, we assume that we have a stable source of squeezed states $|r\rangle$,
as the generation of squeezed states is a mature technology.
A typical squeezing level of squeezed light used for experiments is given by $6.3$ dB, which corresponds to the squeezing parameter
$|r|=0.725$ \cite{Kashiwazaki2021}.
In the current paper, we typically use this parameter for showing
examples but note that the detection
of a $15$ dB squeezed state of light was demonstrated experimentally \cite{Vahlbruch2016}.

In the proposed method, it is critical to adjust the duration of the nonlinear interaction time $\tau$.
To realize a precise adjustment of $\tau$, we can fabricate a cavity of the cross-Kerr nonlinear medium,
which is caught between two half-silvered mirrors.
We can vary $\tau$ by changing the transmittance of mirrors.
To obtain the phase factor $e^{-i2n\kappa\tau}$with $2\kappa\tau=\pi$ for Eq.~(\ref{time-evolution-cross-Kerr-nonlinear-interaction-0}) precisely,
$\kappa$ must be large enough.

\subsection{\label{subsection-generation-entangled-states}Heralded single-photon probability and conditional probability}
Now let us establish
the probability of successfully emitting our heralded single photon.
We input the superposition of oppositely squeezed states $|r;\pm\rangle_{a}$
into
one port of a 50-50 beam splitter with the vacuum on the second
as depicted in Fig.~\ref{figure01}.
The incident squeezed field is transformed to
\begin{eqnarray}
|r;\pm\rangle_{a}|0\rangle_{b}
&\to&
{\cal N}_{\pm}(r)^{-1/2}
[
\hat{S}_{ab}(-\frac{r}{2})\hat{S}_{a}(\frac{r}{2})\hat{S}_{b}(\frac{r}{2}) \nonumber \\
&&
\pm
\hat{S}_{ab}(\frac{r}{2})\hat{S}_{a}(-\frac{r}{2})\hat{S}_{b}(-\frac{r}{2})
]
|0\rangle_{a}|0\rangle_{b},
\label{entangled-states-from-beam-splitter-0}
\end{eqnarray}
where
$\hat{S}_{ab}(\pm r/2)=\exp[(\pm r/2)(\hat{a}^{\dagger}\hat{b}^{\dagger}-\hat{a}\hat{b})]$
are the
origins of the entanglement between the modes.

Next the probability we can detect $n_{a}$ photons at one output port and $n_{b}$ at the other
is given by
\begin{eqnarray}
&&
P(n_{a},n_{b};r;\pm) \nonumber \\
&=&
{\cal N}_{\pm}(r)^{-1}
|
\langle n_{a},n_{b}|\hat{S}_{ab}(-\frac{r}{2})\hat{S}_{a}(\frac{r}{2})\hat{S}_{b}(\frac{r}{2})|0\rangle_{a}|0\rangle_{b} \nonumber \\
&&
\pm
\langle n_{a},n_{b}|\hat{S}_{ab}(\frac{r}{2})\hat{S}_{a}(-\frac{r}{2})\hat{S}_{b}(-\frac{r}{2})|0\rangle_{a}|0\rangle_{b}
|^{2}.
\label{probability-na-nb-superposition-oppositely-squeezed-states-0}
\end{eqnarray}
In contrast, if we input an initial state
$|r\rangle_{a}|0\rangle_{b}=\hat{S}_{a}(r)|0\rangle_{a}|0\rangle_{b}$
onto the beam splitter,
the probability is
\begin{equation}
P(n_{a},n_{b};r)
=
|\langle n_{a},n_{b}|\hat{S}_{ab}(-\frac{r}{2})\hat{S}_{a}(\frac{r}{2})\hat{S}_{b}(\frac{r}{2})|0_{a},0_{b}\rangle|^{2}.
\label{probability-na-nb-squeezed-state-0}
\end{equation}

We can then define our conditional probabilities $P_{\mbox{\scriptsize c}}(r;\pm)$ and $P_{\mbox{\scriptsize c}}(r)$ as
\begin{equation}
P_{\mbox{\scriptsize c}}(r;\pm)
=
P(1,1;r;\pm)/\sum_{n=0}^{\infty}P(1,n;r;\pm).
\label{definition-efficiency-r-pm-0}
\end{equation}
\begin{equation}
P_{\mbox{\scriptsize c}}(r)
=
P(1,1;r)/\sum_{n=0}^{\infty}P(1,n;r).
\end{equation}
If we observe $|\pm\alpha\rangle_{2}$ with the homodyne measurement,
inject $|r;\pm\rangle_{a}$ into the beam splitter,
and detect the heralding single photon in one port,
we then obtain a single photon from the other port with probabilities $P_{\mbox{\scriptsize c}}(r;\pm)$.
Thus, we can regard $P_{\mbox{\scriptsize c}}(r;\pm)$ as the conditional probabilities of single-photon emission
after the postselection with the measurement outcomes on the other port.
Similarly
if we inject $|r\rangle_{a}$ into the beam splitter and detect the heralding single photon,
we obtain a single photon with conditional probability $P_{\mbox{\scriptsize c}}(r)$.

\subsection{\label{subsection-numerical-calculations}Ideal single photon detection}
The photon after the beam splitter with $(n_{a},n_{b})=(1,1)$ can be considered as a heralded single photon.
When we generate $|r;\pm\rangle_{1}$ from $|r\rangle_{1}$,
the probability that we obtain $|r;\pm\rangle_{1}$ is given by ${\cal N}_{\pm}(r)/4$.
Thus, the actual probability for obtaining $n_{a}$ and $n_{b}$ photons from $|r;\pm\rangle_{1}$ is given by
${\cal N}_{\pm}(r)P(n_{a},n_{b};r;\pm)/4$.

\begin{figure}
\begin{center}
\includegraphics[width=0.9\linewidth]{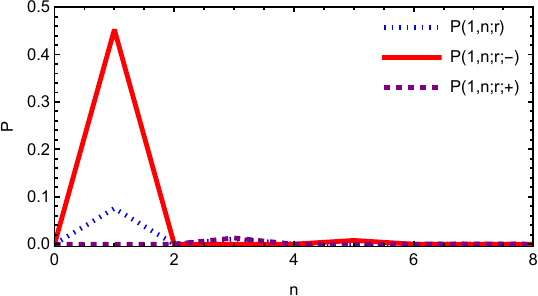}
\end{center}
\caption{Plot of the probability $P(1,n;r)$ and $P(1,n;r;\pm)$. We generate an $n$ photon state $|n\rangle$ in one mode and a heralding photon in the other for the 
injected states $|r\rangle_{a}$ and $|r;\pm\rangle_{a}$
with
$r=0.725$.
The dotted blue, solid red, and dashed purple line graphs represent $P(1,n;r)$, $P(1,n;r;-)$, and $P(1,n;r;+)$, respectively.
The line graph of $P(1,n;r;+)$ overlaps the horizontal axis of $n$ mostly.}
\label{figure02}
\end{figure}

\begin{figure}
\begin{center}
\includegraphics[width=\linewidth]{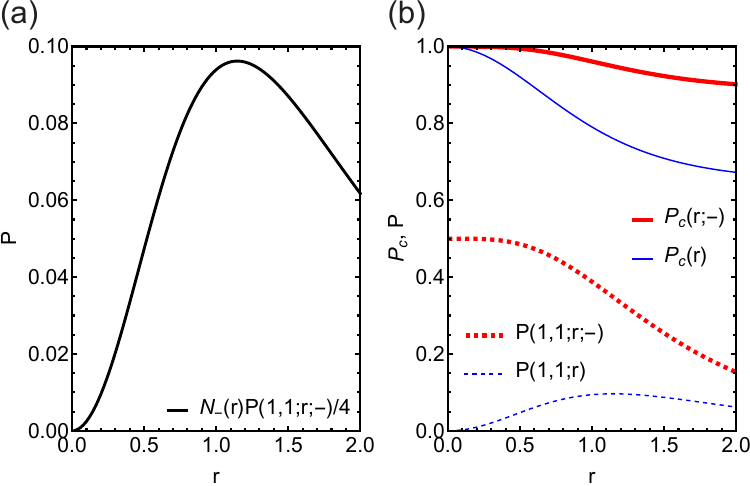}
\end{center}
\caption{(a) Plot of the probability
${\cal N}_{-}(r)P(1,1;r;-)/4$.
We obtain a heralded single photon
by injecting $|r;-\rangle_{a}$
for $0\leq r\leq 2$.
When $r=1.146$, the graph reaches its maximum value $0.096{\,}23$.
Thus, the optimal value of $r$ to maximize the probability is $r=1.146$.
(b) Plot of the probabilities of the heralded single photon as functions of $r$.
The thick solid red and thin solid blue curves represent $P_{\mbox{\scriptsize c}}(r;-)$ and $P_{\mbox{\scriptsize c}}(r)$, respectively.
The thick dashed red and thin dashed blue curves represent $P(1,1;r;-)$ and $P(1,1;r)$, respectively.}
\label{figure03}
\end{figure}

In Fig.~\ref{figure02}, we plot the probability $P(1,n;r)$ and $P(1,n;r;\pm)$ that we generate an $n$ photon state $|n\rangle$ in one mode
and a heralding photon in the other for the input states
$|r\rangle_{a}$ and $|r;\pm\rangle_{a}$
with
$r=0.725$.
Looking at Fig.~\ref{figure02}, we note that
$P(1,1;r;-)=0.453$,
$P(1,5;r;-)=7.85\times 10^{-3}$,
and 
$P(1,n;r;-)=0$ for $n=0,2,3,4$.
Moreover, we obtain $P_{\mbox{\scriptsize c}}(r;-)=0.983$.
If we generate the superposition of the oppositely squeezed states $|r;-\rangle_{a}$,
it
should allow us to realize
the heralded single-photon source with a high
probability and so be
a good alternative to the SPDC.
By contrast, if we prepare the squeezed state $|r\rangle_{a}$,
we obtain $P(1,1;r)=7.54\times 10^{-2}$ with $P_{\mbox{\scriptsize c}}(r)=0.859$.
Looking at the plot of $|r;-\rangle$, the following question comes to mind.
There must be an optimal value of $r$ to maximize the probability.
Answers to this question are shown in Fig.~\ref{figure03}.

\begin{figure}
\begin{center}
\includegraphics[width=\linewidth]{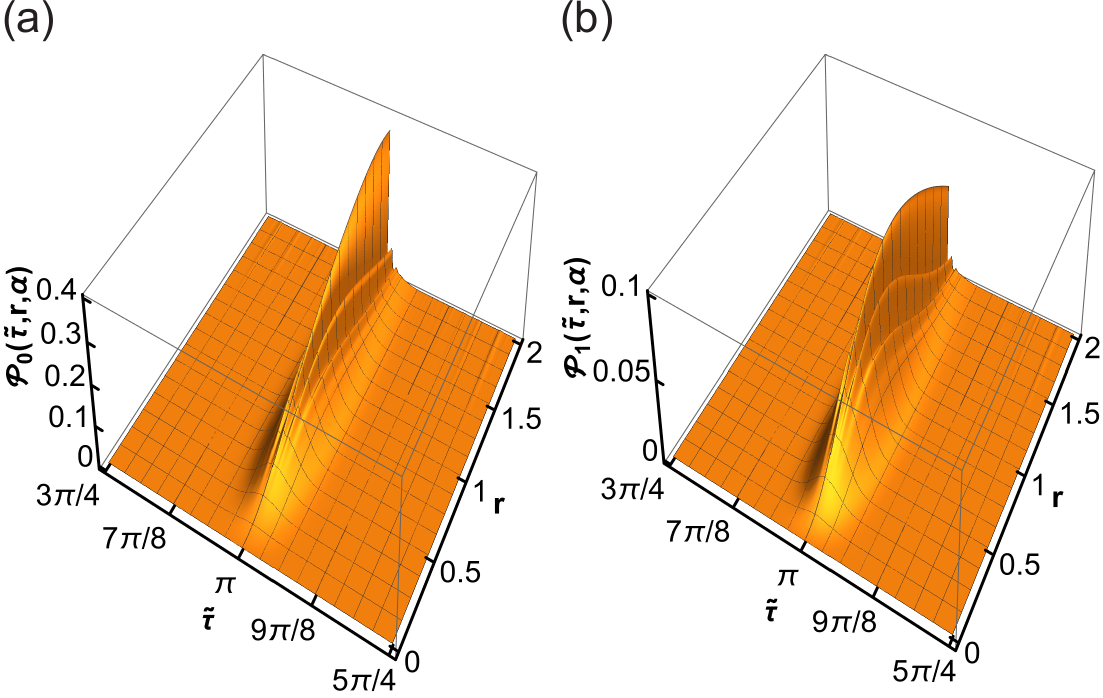}
\end{center}
\caption{(a) Plot of ${\cal P}_{0}(\tilde{\tau},r,\alpha)$ as a function of $\tilde{\tau}$ and $r$ with $\alpha=10$.
(b) Plot of ${\cal P}_{1}(\tilde{\tau},r,\alpha)$ as a function of $\tilde{\tau}$ and $r$ with $\alpha=10$.}
\label{figure04}
\end{figure}

Next in Fig.~\ref{figure03}(a), we plot the actual probability that we obtain
the $|1\rangle$ photon Fock state
conditioned that we observe a heralding single photon for the initial state $|r;-\rangle_{a}$.
Examining Fig.~\ref{figure03}(a), we note that ${\cal N}_{-}(r)P(1,1;r;-)/4$ reaches its maximum value of $0.096{\,}23$
when $r=1.146$.
At that point, the conditional probability is $P_{\mbox{\scriptsize c}}(1.146;-)=0.9488$.
Figure~\ref{figure03}(b) shows plots of the probabilities of the heralded single photon as a function of $r$.
We observe that $P_{\mbox{\scriptsize c}}(r;-)$
is
larger than or equal to 
$P_{\mbox{\scriptsize c}}(r)$ for any $r$.
Thus, as the heralded single-photon source,
the superposition of oppositely squeezed states $|r;-\rangle$ is preferable to the typical squeezed state.
It is useful to mention that the
squeezed state $|r\rangle$ is a superposition of $\{|2n\rangle: n=0,1,2,...\}$
while contrastingly, the superpositions of the oppositely squeezed states $|r;+\rangle$ and $|r;-\rangle$ are superpositions of
$\{|2m\rangle: m=0,2,4,...\}$ and $\{|2m\rangle: m=1,3,5,...\}$,
respectively.
The Fock states present in these superpositions of squeezed states are further separated in photon number space than the usual squeezed state.
If we can prepare
$|r;-\rangle\propto c_{2}|2\rangle+c_{6}|6\rangle+...$
and inject that state into the 50-50 beam splitter,
a transformation
$|2\rangle_{a}|0\rangle_{b}
\to
(1/2)(|2\rangle_{a}|0\rangle_{b}+|0\rangle_{a}|2\rangle_{b})
-
(1/\sqrt{2})|1\rangle_{a}|1\rangle_{b}$
occurs, and
we obtain a state $|1\rangle_{a}|1\rangle_{b}$ with
a maximal probability of $1/2$.
Hence, we can implement the heralded single-photon
source with higher conditional probability.
Figures~\ref{figure04}(a) and \ref{figure04}(b) show 3D plots of the probabilities of generating $|r;-\rangle$ and a heralded single photon
as functions of $\tilde{\tau}=2\kappa\tau$ and $r$, respectively,
with fixing $\alpha$ at a specific value.
In Fig.~\ref{figure04}(a), we plot ${\cal P}_{0}$, the probability of generating $|r;-\rangle$, where
\begin{equation}
{\cal P}_{0}(\tilde{\tau},r,\alpha)
=
|_{12}
\langle\Psi_{\mbox{\scriptsize out}}|r;-\rangle_{1}|-\alpha\rangle_{2}
|^{2}.
\end{equation}
In Fig.~\ref{figure04}(b), we plot ${\cal P}_{1}$, the probability of generating a heralded single photon, where
\begin{equation}
{\cal P}_{1}(\tilde{\tau},r,\alpha)
=
P(1,1,r;-){\cal P}_{0}(\tilde{\tau},r,\alpha).
\end{equation}
Looking at both plots, we observe sharp peaks at $\tilde{\tau}=\pi$.

\subsection{\label{subsection-errors}Imperfections and errors}
\begin{figure}
\begin{center}
\includegraphics[width=\linewidth]{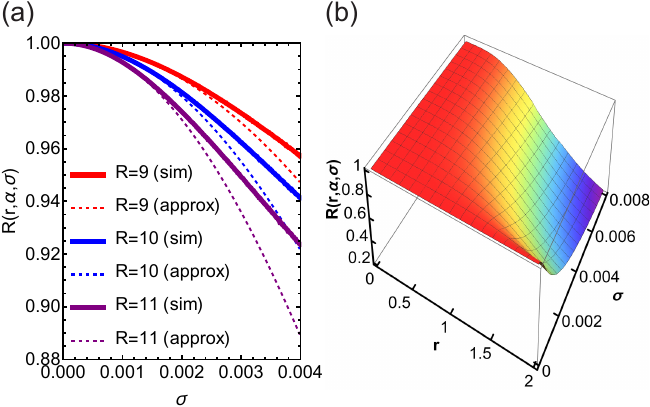}
\end{center}
\caption{(a) Plot of $R(r,\alpha,\sigma)$ as a function of $\sigma$ for $r=0.725$ and $\alpha=9$ (thick solid red line),
$10$ (thick dashed blue line), and $11$ (thick dotted purple lines), respectively.
The thin solid red, thin dashed blue, and thin dotted purple curves represent functions $\exp(-\lambda\sigma^{2})$ for $\lambda=3401\pm 3$, $5102\pm 2$, and $7360\pm 4$, respectively,
that
approximate the curves of $R(r,\alpha,\sigma)$ well for small $r$.
(b) 3D plot of $R(r,\alpha,\sigma)$ as a function of both $r$ and $\sigma$ for $\alpha=10$.
In the regime where $r$ is larger than $1/2$, $R(r,\alpha,\sigma)$ decreases as $\sigma$ gets larger.}
\label{figure05}
\end{figure}

In any realistic implementation, there will be imperfections, and so
let us explore effects associated with the cross-phase modulation and also single-photon heralding detection.
We have previously detailed
the generation of $|r;-\rangle$ from the squeezed state $|r\rangle$ using the nonlinear cross-Kerr medium.
For this generation,
we had set duration $\tau$ for the nonlinear interaction as $2\kappa\tau=\pi$.
Here, we consider a case where an inaccuracy occurs;
such timing is not
perfect and a phase error could be induced as
$2\kappa\tau=\pi+\Delta\theta$.
In this case
\begin{eqnarray}
|\Psi'_{\mbox{\scriptsize out}}\rangle_{12}
&=&
\frac{1}{\sqrt{\cosh r}}
\sum_{n=0}^{\infty}
\frac{\sqrt{(2n)!}}{2^{n}n!}(-\tanh r)^{n} \nonumber \\
&&
\times
|2n\rangle_{1}|(-1)^{n}\alpha e^{-in\Delta\theta}\rangle_{2},
\end{eqnarray}
where we assume $\alpha$ is real and greater than zero.
To realize our heralded single-photon source,
we want to generate a state of the form
\begin{equation}
|\Psi^{(-)}_{\mbox{\scriptsize out}}\rangle_{12}
=
\frac{1}{2}
{\cal N}_{-}(r)^{1/2}|r;-\rangle_{1}|-\alpha\rangle_{2},
\end{equation}
and so to evaluate the effects caused by $\Delta\theta$,
we define the following ratio
that implies the decrease in generation probability of $|r;-\rangle$ due to $\Delta\theta$:
\begin{equation}
R(r,\alpha,\Delta\theta)
=
\frac
{|_{12}\langle\Psi^{(-)}_{\mbox{\scriptsize out}}|\Psi'_{\mbox{\scriptsize out}}\rangle_{12}|^{2}}
{\left.
|_{12}\langle\Psi^{(-)}_{\mbox{\scriptsize out}}|\Psi'_{\mbox{\scriptsize out}}\rangle_{12}|^{2}
\right|_{\Delta\theta=0}
}.
\label{definition-Ratio-0}
\end{equation}
Here
$\Delta\theta$ is not a constant value but a probabilistic
random variable that
obeys Gaussian distribution whose mean value and standard deviation are given by zero and $\sigma$, respectively.
Taking the average of $R(r,\alpha,\Delta\theta)$ with varying $\Delta\theta$ for many samples,
we numerically obtain
\begin{equation}
R(r,\alpha,\sigma)
=
\int_{-\infty}^{\infty}d(\Delta\theta)
\frac{1}{\sqrt{2\pi\sigma^{2}}}
\exp(-\frac{\Delta\theta^{2}}{2\sigma^{2}})R(r,\alpha,\Delta\theta).
\end{equation}
In Fig.~\ref{figure05}(a), we plot $R(r,\alpha,\sigma)$ for $\alpha=9$, $10$, and $11$ with $r=0.725$.
In the range of $0\leq\sigma\leq 0.001$,
the function $\exp(-\lambda\sigma^{2})$ approximates to the curves of $R(r,\alpha,\sigma)$ well.
Looking at those plots, we note that the curve of $R(r,\alpha,\sigma)$ gets closer to a horizontal
as
$\alpha$ becomes smaller.
Thus, we become aware that the error of the inaccuracy of the duration is serious for large $\alpha$.
Because the homodyne measurement requires a strong coherent state,
we need to increase $|\alpha|$ for precise measurements.
However, large $|\alpha|$ makes the single-photon emission vulnerable to the inaccuracy of duration $\tau$, and a trade-off emerges.
Figure~\ref{figure05}(b) shows a 3D plot of $R(r,\alpha,\sigma)$ with $\alpha=10$ as a function of $r$ and $\sigma$
where we
note that, in the range where $r$ is larger than $1/2$,
$R(r,10,\sigma)$ decreases as $\sigma$ gets larger.
Fluctuations
of the phase error
are problematic
when we try to obtain strong squeezing.
If we want to generate a strongly squeezed state, we must ensure the accuracy of the cross-Kerr phase modulation
giving us
another trade-off.

In Fig.~\ref{figure05}(a), $R(r,\alpha,\sigma)$ departs from unity considerably for $\sigma=0.004$.
This standard deviation corresponds to a fluctuation $\Delta\tau=\tau\sigma/\pi$.
This evaluation suggests to us that we must fix $\tau$ with the accuracy $\Delta\tau/\tau\sim 10^{-3}$,
and it is well within our reach experimentally.

Next, we consider inefficient single-photon detection.
So far, we have assumed that we can use an ideal single-photon detector for
the heralding measurement.
The implementation of the single-photon detector is a cutting-edge technology
with detection efficiencies above 90{\%} regularly reported \cite{Yuan2007,Jiang2007}.

We can model a ``click/no click'' detector
of efficiency $\eta$ by
the POVM $\{\hat{M},\hat{\mbox{\boldmath $I$}}-\hat{M}\}$
\cite{Kok2007,Resch2001,Akhlaghi2011},
where
\begin{equation}
\hat{M}
=
\eta\sum_{k=1}^{\infty}(1-\eta)^{k-1}|k\rangle\langle k|,
\label{definition-click-operator-0}
\end{equation}
for a ``click''
and $\hat{\mbox{\boldmath $I$}}-\hat{M}$
for ``no click''.
If we use our quantum state $|r;-\rangle$ for a heralded single-photon source,
the click probability is then
\begin{equation}
P_{\mbox{\scriptsize click}}(r,\eta;-)
=
\eta
\sum_{n_{a}=1}^{\infty}
\sum_{m_{b}=0}^{\infty}
(1-\eta)^{n_{a}-1}
P(n_{a},m_{b};r;-).
\end{equation}
The probability that the detector
``clicks''
and the heralded single photon is actually emitted is given by
\begin{equation}
P_{\mbox{\scriptsize click},1}(r,\eta;-)
=
\eta
\sum_{n_{a}=1}^{\infty}
(1-\eta)^{n_{a}-1}
P(n_{a},1;r;-).
\end{equation}
Then, we can compute the conditional probability as
\begin{equation}
P_{\mbox{\scriptsize click,c}}(r,\eta;-)
=
\frac{P_{\mbox{\scriptsize click},1}(r,\eta;-)}{P_{\mbox{\scriptsize click}}(r,\eta;-)}.
\end{equation}

\begin{figure}
\begin{center}
\includegraphics[width=\linewidth]{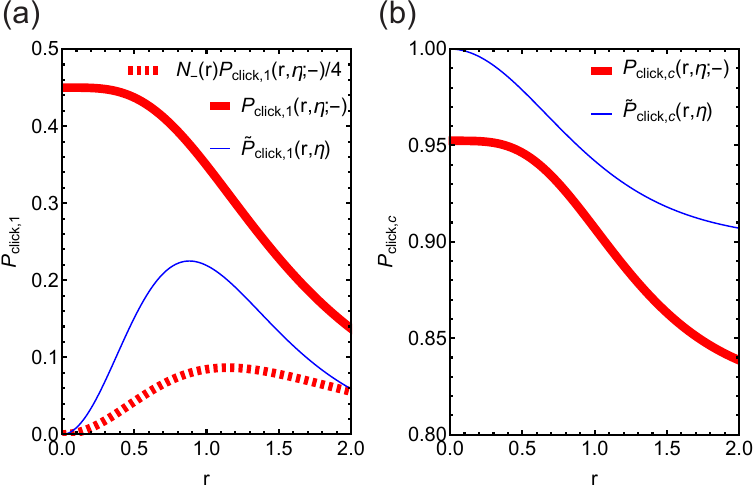}
\end{center}
\caption{(a) Plot of $P_{\mbox{\scriptsize click},1}(r,\eta;-)$, $\tilde{P}_{\mbox{\scriptsize click},1}(r,\eta)$
defined in Eq.~(\ref{tildeQ1-definition0}),
and ${\cal N}_{-}(r)P_{\mbox{\scriptsize click},1}(r,\eta;-)/4$ as a function of $r$ for $\eta=0.9$,
where they are represented by the thick solid red, thin solid blue, and thick dashed red curves, respectively.
(b) Plot of $P_{\mbox{\scriptsize click,c}}(r,\eta;-)$ and $\tilde{P}_{\mbox{\scriptsize click,c}}(r,\eta)$
defined in Eq.~(\ref{tildeE-definition0})
as a function of $r$ with $\eta=0.9$,
where they are represented by the thick red and thin blue curves, respectively.}
\label{figure06}
\end{figure}

\begin{figure}
\begin{center}
\includegraphics[width=\linewidth]{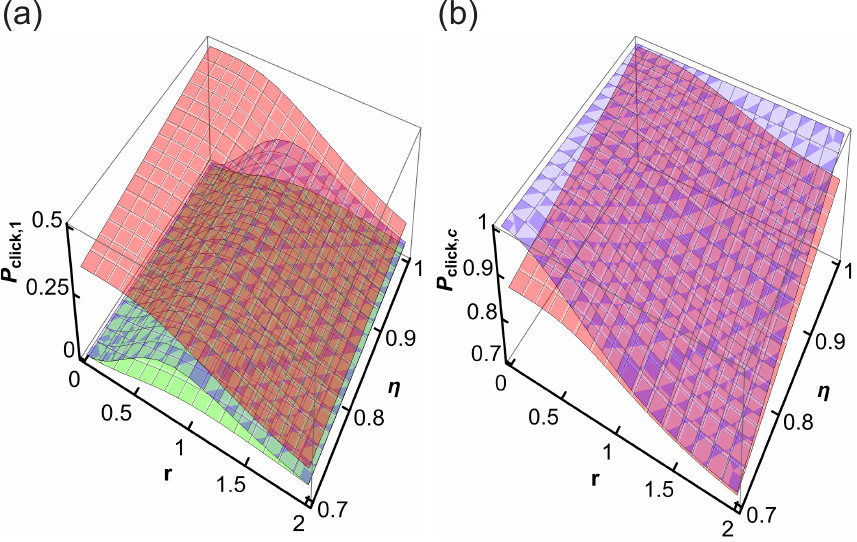}
\end{center}
\caption{(a) 3D plots of $P_{\mbox{\scriptsize click},1}(r,\eta;-)$, $\tilde{P}_{\mbox{\scriptsize click},1}(r,\eta)$, and ${\cal N}_{-}(r)P_{\mbox{\scriptsize click},1}(r,\eta;-)/4$ as functions of $r$ and $\eta$,
where they are represented by the curved red, blue, and green transparent surfaces, respectively.
(b) 3D plots of $P_{\mbox{\scriptsize click,c}}(r,\eta;-)$ and $\tilde{P}_{\mbox{\scriptsize click,c}}(r,\eta)$ as functions of $r$ and $\eta$,
where they are represented by the curved red and blue transparent surfaces, respectively.
In the regime $r\leq 2$ and $0.7 \leq \eta \leq 1$ we have
$P_{\mbox{\scriptsize click},1}(r,\eta;-)>\tilde{P}_{\mbox{\scriptsize click},1}(r,\eta)>{\cal N}_{-}(r)P_{\mbox{\scriptsize click},1}(r,\eta;-)/4$
and
$P_{\mbox{\scriptsize click,c}}(r,\eta;-)<\tilde{P}_{\mbox{\scriptsize click,c}}(r,\eta)$.}
\label{figure07}
\end{figure}

In Fig.~\ref{figure06}(a), we plot $P_{\mbox{\scriptsize click},1}(r,\eta;-)$ and ${\cal N}_{-}(r)P_{\mbox{\scriptsize click},1}(r,\eta;-)/4$ as functions of $r$ for $\eta=0.9$,
where ${\cal N}_{-}(r)P_{\mbox{\scriptsize click},1}(r,\eta;-)/4$ is the probability that the heralded single photon is actually emitted.
Although the single-photon detector is
imperfect ($\eta=0.9$),
$P_{\mbox{\scriptsize click},1}(r,\eta;-)$ is nearly equal to $0.5$ for the small squeezing parameter $r$.
This is one of the advantages
that our method has.

In Fig.~\ref{figure06}(b), we plot $P_{\mbox{\scriptsize click,c}}(r,\eta;-)$ as a function of $r$ for $\eta=0.9$.
This plot shows that $P_{\mbox{\scriptsize click,c}}(r,\eta;-)$ is nearly equal to $0.95$ at $r=0$ and decreases gradually as $r$ becomes larger.
In Fig.~\ref{figure07}(a), we plot $P_{\mbox{\scriptsize click},1}(r,\eta;-)$ and ${\cal N}_{-}(r)P_{\mbox{\scriptsize click},1}(r,\eta;-)/4$ as functions of $r$ and $\eta$,
where ${\cal N}_{-}(r)P_{\mbox{\scriptsize click},1}(r,\eta;-)/4$ is the probability that the heralded single photon is actually emitted.
In the limit of $r\to 0$, $P_{\mbox{\scriptsize click},1}(r,\eta;-)$ is finite and nonzero, but ${\cal N}_{-}(r)P_{\mbox{\scriptsize click},1}(r,\eta;-)/4$ approaches zero.
In Fig.~\ref{figure07}(b), we plot $P_{\mbox{\scriptsize click,c}}(r,\eta;-)$
as a function of $r$ and $\eta$.
As $r$ increases and $\eta$ decreases, $P_{\mbox{\scriptsize click,c}}(r,\eta;-)$ becomes smaller gradually.

\section{\label{section-second-method}Comparison with a heralded single-photon source from the two-mode squeezed state}
Let us now review the generation of a heralded single photon using the pure two-mode squeezed state for comparison with our proposed method.
Injecting two squeezed states individually onto
a 50-50 beam splitter
allows us to create the two-mode squeezed state
\begin{equation}
|r\rangle_{a}|r\rangle_{b}
\to
\hat{S}_{ab}(ir)|0\rangle_{a}|0\rangle_{b},
\label{rr-beam-splittre-transform-0}
\end{equation}
where $r$ is real.
The probability we detect $n_{a}$ photons and $n_{b}$ photons
is given by
\begin{equation}
\tilde{P}(n_{a},n_{b};r)
=
\left|
\langle n_{a},n_{b}|\hat{S}_{ab}(ir)|0_{a},0_{b}\rangle
\right|^{2},
\end{equation}
meaning that we can write the probability for
$n_{a}=1$ as
$
\tilde{P}(1,n_{b};r)
=
(\tanh r/\cosh r)^{2}
\delta_{n_{b},1}
$.
If we observe a single photon in the first output mode,
we obtain a single photon in the second output mode deterministically.

\begin{figure}
\begin{center}
\includegraphics[width=0.9\linewidth]{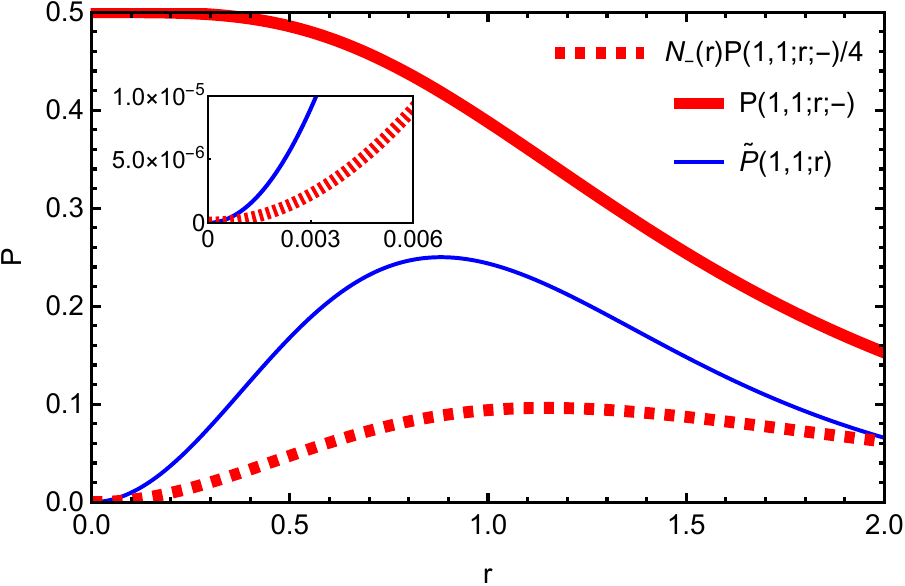}
\end{center}
\caption{Plot of $P(1,1;r;-)$, $\tilde{P}(1,1;r)$, and ${\cal N}_{-}(r)P(1,1;r;-)/4$ as functions of $r$
which are represented as the thick solid red, thin solid blue, and thick dashed red curves, respectively.
The graph of $\tilde{P}(1,1;r)$ reaches the maximum value $1/4$ for $r=\mbox{arcsech}(1/\sqrt{2})=0.881$
and reaches the minimum value $0$ for $r=0$.
We put a small frame of plots of ${\cal N}_{-}(r)P(1,1;r;-)/4$ and $\tilde{P}(1,1;r)$ for $0\leq r\leq 0.006$ in the main frame.
These plots show that ${\cal N}_{-}(r)P(1,1;r;-)/4$ and $\tilde{P}(1,1;r)$ are larger than $4\times 10^{-6}$ for $r\geq 0.004$.
Thus, the heralded single-photon sources of our scheme and the pure two-mode squeezed state are
more efficient than the SPDC even for small squeezing parameter $r$.
}
\label{figure08}
\end{figure}

In Fig.~\ref{figure08}, we plot $P(1,1;r;-)$, $\tilde{P}(1,1;r)$, and ${\cal N}_{-}(r)P(1,1;r;-)/4$
as a function of $r$.
When $r=\mbox{\rm arcsech}(1/\sqrt{2})=0.881$,
$\tilde{P}(1,1;r)$ achieves its maximum value of $1/4$.
In the limit of $r\to 0$, $P(1,1;r;-)$ and $\tilde{P}(1,1;r)$ approach $1/2$ and $0$, respectively.
In the limit of $r\to 0$,
both ${\cal N}_{-}(r)P(1,1;r;-)/4$ and $\tilde{P}(1,1;r)$ approach zero,
sharing the same defect.
However, in a practical circumstance where the squeezing parameter $r$ is small,
our scheme with the postselection of measuring $|-\alpha\rangle_{2}$ shown in Fig.~\ref{figure01}
gives the probability $P(1,1;r;-)$ to be nearly equal to $1/2$ around $r=0$.
Hence, the superposition of oppositely squeezed states $|r;-\rangle$ is preferable to the pure two-mode squeezed state $\hat{S}_{ab}(ir)|0\rangle_{a}|0\rangle_{b}$
as a source of the heralded single photons.

Now if we prepare a factory where we can produce many light beams of $|r;-\rangle$ by the cross-Kerr nonlinear medium
and the homodyne measurements,
we obtain heralded single photons efficiently for small $r$
because of $\lim_{r \to 0}P(1,1;r;-)=0.5$.
Moreover, looking at Fig.~\ref{figure03}(b), we note that the conditional probability $P_{\mbox{\scriptsize c}}(r;-)$ is close to unity for small $r$.
Hence, if we construct the heralded single-photon source with $|r;-\rangle$,
the emission of the single photons is nearly deterministic.
Contrastingly, if we use the pure two-mode squeezed state $\hat{S}_{ab}(ir)|0\rangle_{a}|0\rangle_{b}$
for the heralded single-photon source,
we cannot obtain the emission of the single photons efficiently because $\lim_{r \to 0}\tilde{P}(1,1;r)=0$.
In Fig.~\ref{figure08}, we note that $\tilde{P}(1,1;r)<P(1,1;r;-)$ for $0\leq r\leq 2$.
Thus, $|r;-\rangle_{a}$ is preferable to $\hat{S}_{ab}(ir)|0\rangle_{a}|0\rangle_{b}$.
Although
we always obtain the heralded single photon when we detect the heralding single photon
for the two-mode squeezed state
and its conditional probability is exactly equal to unity,
looking at Fig.~\ref{figure03}(b),
we observe that $P_{\mbox{\scriptsize c}}(r;-)$ is nearly and sufficiently equal to unity, as well.
Thus, regarding the conditional probabilities, the superposition of oppositely squeezed states $|r;-\rangle$ is as useful as the pure two-mode squeezed state.

Next, we consider the case where the single-photon detector is imperfect, and its ``click'' operator is given by Eq.~(\ref{definition-click-operator-0}).
In this case, the click probability is
\begin{equation}
\tilde{P}_{\mbox{\scriptsize click}}(r,\eta)
=
\frac{2\eta \tanh^{2}r}{2-\eta[1-\cosh(2r)]}.
\end{equation}
The probability that the detector observes the click operator $\hat{M}$ and the heralded single photon is actually emitted is given by
\begin{equation}
\tilde{P}_{\mbox{\scriptsize click},1}(r,\eta)
=
\eta
\frac{\tanh^{2}r}{\cosh^{2}r}.
\label{tildeQ1-definition0}
\end{equation}
Then, we can compute the conditional probability as
\begin{equation}
\tilde{P}_{\mbox{\scriptsize click,c}}(r,\eta)
=
\frac{\tilde{P}_{\mbox{\scriptsize click},1}(r,\eta)}{\tilde{P}_{\mbox{\scriptsize click}}(r,\eta)}
=
\eta+(1-\eta)\cosh^{-2}r.
\label{tildeE-definition0}
\end{equation}

In Figs.~\ref{figure06}(a) and \ref{figure06}(b), we plot $\tilde{P}_{\mbox{\scriptsize click},1}(r,\eta)$ and $\tilde{P}_{\mbox{\scriptsize click,c}}(r,\eta)$ as functions of $r$ for $\eta=0.9$
while in
Figs.~\ref{figure07}(a) and \ref{figure07}(b), we plot $\tilde{P}_{\mbox{\scriptsize click},1}(r,\eta)$ and $\tilde{P}_{\mbox{\scriptsize click,c}}(r,\eta)$ as functions of $r$ and $\eta$.
Figure~\ref{figure06}(a) shows that $\tilde{P}_{\mbox{\scriptsize click},1}(r,\eta)$ is nearly equal to zero for small $r$.
This characteristic is in contrast to our proposed method
because $P_{\mbox{\scriptsize click},1}(r,\eta;-)$ is larger than $0.4$ for small $r$.
Comparing $P_{\mbox{\scriptsize click,c}}(r,\eta;-)$ and $\tilde{P}_{\mbox{\scriptsize click,c}}(r,\eta)$ in Fig.~\ref{figure06}(b), we cannot find a large difference between them for small $r$.
For this point, both methods are comparable.
Examining Fig.~\ref{figure07}(a), we note that $P_{\mbox{\scriptsize click},1}(r,\eta;-)$ is larger than $\tilde{P}_{\mbox{\scriptsize click},1}(r,\eta)$
for $0\leq r\leq 2$ and $0.7\leq\eta\leq 1$.
Moreover, $\lim_{r\to 0}P_{\mbox{\scriptsize click},1}(r,\eta;-)>0$ and $\lim_{r \to 0}\tilde{P}_{\mbox{\scriptsize click},1}(r,\eta)=0$ hold.
Thus, for most realistic cases,
the superposition of oppositely squeezed states is more beneficial to the heralded single-photon emission
than the pure two-mode squeezed state.
Looking at Fig.~\ref{figure07}(b), we note that $P_{\mbox{\scriptsize click,c}}(r,\eta;-)<\tilde{P}_{\mbox{\scriptsize click,c}}(r,\eta)$
for $0\leq r\leq 2$ and $0.7\leq \eta\leq 1$.
However,
for larger $r$ and smaller $\eta$ the difference is small.

\section{\label{section-second-order-intensity-correlation-functions}The second-order intensity correlation functions}
Now, we consider the second-order intensity correlation functions $g^{(2)}(0)$
for the heralded single-photon emitters realized with the superposition of oppositely squeezed states
$|r;-\rangle$
and the pure two-mode squeezed state $\hat{S}_{ab}(ir)|0\rangle_{a}|0\rangle_{b}$.
We can measure the quality of the single-photon source by
\begin{equation}
g^{(2)}(0)
=
\frac{\langle n(n-1)\rangle}{\langle n\rangle^{2}}.
\end{equation}
When an on-demand identical photon gun is realized,
we obtain $g^{(2)}(0)=0$.
For our proposed method, we have to calculate $g^{(2)}(0)$ numerically.
By contrast, for the pure two-mode squeezed state,
\begin{equation}
g^{(2)}(0)
=
-3
+\frac{2}{\eta}
+\eta
+(1-\eta)\cosh 2r.
\end{equation}

\begin{figure}
\begin{center}
\includegraphics[width=\linewidth]{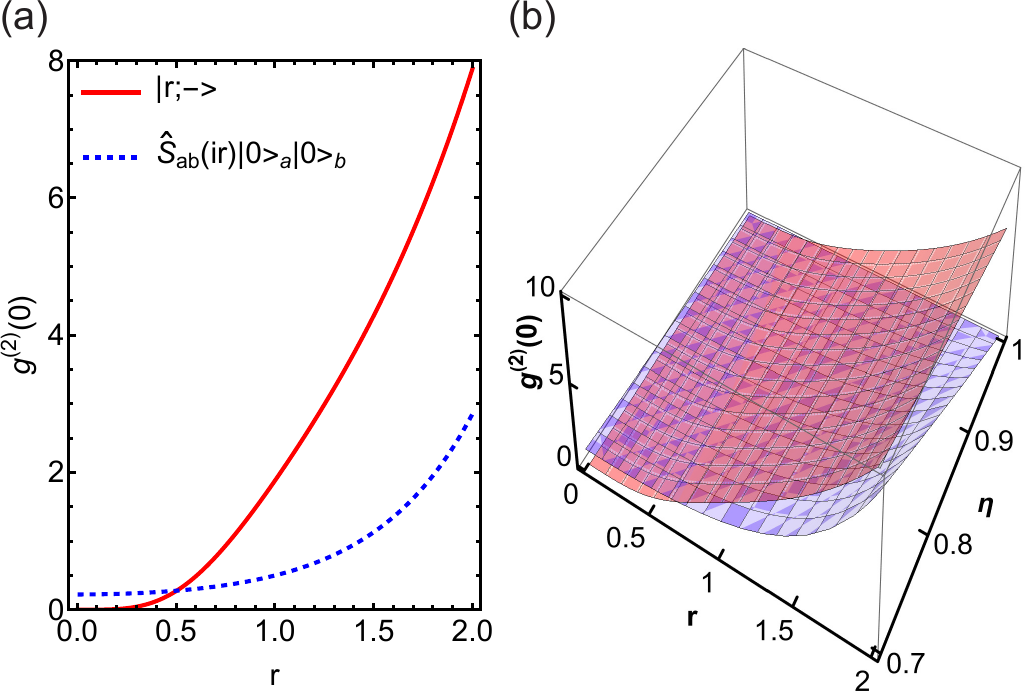}
\end{center}
\caption{(a) Plot of $g^{(2)}(0)$ for $|r;-\rangle$ (solid red curve) and $\hat{S}_{ab}(ir)|0\rangle_{a}|0\rangle_{b}$
(dashed blue curve) as a function of $r$ with $\eta=0.9$.
(b) Plot of $g^{(2)}(0)$ as a function of $r$ and $\eta$.}
\label{figure09}
\end{figure}

We plot $g^{(2)}(0)$ for $|r;-\rangle$
and $\hat{S}_{ab}(ir)|0\rangle_{a}|0\rangle_{b}$ in Fig.~\ref{figure09}.
We observe that $g^{(2)}(0)$ is always less for $|r;-\rangle$ in the shown region.
In particular for $|r;-\rangle$, we notice that $g^{(2)}(0)$ approaches zero as $r\to 0$,
while for the two-mode squeezed state it is always greater than zero for $r\geq 0$.
Thus, for $\eta=0.9$ and $0\leq r\leq 0.504$,
the single-photon emitter realized by $|r;-\rangle$ has a better quality than
the two-mode squeezed state.
We can understand the reason why this difference emerges as follows.
The state $|r;-\rangle$ is a superposition of photon-number states,
$|2\rangle$, $|6\rangle$, $|10\rangle$, ....
Thus, dividing the state $|r;-\rangle$
on two ports of a
50-50 beam splitter,
the probability that the imperfect photon detector detects the state $|2\rangle_{a}$ is suppressed.
Thus, $g^{(2)}(0)$ hardly suffers from an imperfection caused by the term $\eta(1-\eta)|2\rangle_{a}{}_{a}\langle 2|$
included in the observable $\hat{M}$ given by Eq.~(\ref{definition-click-operator-0}).
By contrast, the pure two-mode squeezed state is a superposition of the states,
$|0\rangle_{a}|0\rangle_{b}$,
$|1\rangle_{a}|1\rangle_{b}$,
$|2\rangle_{a}|2\rangle_{b}$,
...,
so that it is vulnerable to the imperfect term $\eta(1-\eta)|2\rangle_{a}{}_{a}\langle 2|$.
In Fig.~\ref{figure09}(b), we plot $g^{(2)}(0)$ as a function of $r$ and $\eta$
for $0\leq r\leq 2$ and $0.7\leq \eta\leq 1$.
We observe that $g^{(2)}(0)$
for our odd superposition of squeezed states is always smaller than that associated with the two-mode squeezed state
for small $r$.
This fact implies that the heralded single-photon emitter realized by $|r;-\rangle$ is more robust than that realized by
$\hat{S}_{ab}(ir)|0\rangle_{a}|0\rangle_{b}$
under the condition that we must use the imperfect single-photon detector.
In Fig.~\ref{figure09}(a),
we note that $g^{(2)}(0)$ of the pure two-mode squeezed state exhibits partial stability around $g^{(2)}(0)=0.22$ for $0\leq r\leq 1.0$.
In contrast, our proposed method can achieve a high quality of single photons
for the small $r$ regime ($0\leq r\leq 0.504$).
This regime of the squeezing parameter $r$ is most relevant in the current experiment.

To compare our scheme with other single-photon sources,
the method of the pure two-mode squeezed state in particular,
we also need to count the probability of single-photon generation as shown in Fig.~\ref{figure03}.
When $r$ is smaller than $0.504$, the success probability of obtaining a single photon after the postselection is nearly equal to $1/2$,
as shown by $P(1,1;r;-)$ in Fig.~\ref{figure03}(b),
hence both the success probability and the quality are high enough.
However, as Fig.~\ref{figure03}(a) indicates,
the yielding of the success case is rather low when $r$ approaches zero.
The usual trade-off between purity and overall efficiency remains in our scheme \cite{Christ2012,Senellart2017}.

\section{\label{section-discussion}Conclusion}
In this paper, we
proposed a method to
generate a heralded single-photon source created from a superposition of oppositely squeezed states.
We outlined the method for producing this state by injecting squeezed and coherent light beams onto a cross-Kerr nonlinear medium
and estimated the probability and conditional probability of the emission of the heralded single photons.
To evaluate the quality of our proposed method,
we used the pure two-mode squeezed state for comparison.
When the single-photon detector detects the heralding signals inaccurately,
our method shows better performances for the second-order intensity correlation function $g^{(2)}(0)$
than the method of the pure two-mode squeezed states
for small $r$ as explained in Sec.~\ref{section-second-order-intensity-correlation-functions}.
This is because $|r\rangle-|-r\rangle$ is a superposition of photon-number states,
$|2\rangle$,
$|6\rangle$,
$|10\rangle$,
...
and
the probability amplitude of $|n\rangle_{a}|2\rangle_{b}$ for $n=0,1,2,...$ is suppressed
when we inject it into the 50-50 beam splitter.
That is an important advantage of our scheme.
However the generation of the odd superposition of squeezed states is more difficult leading to a natural trade-off.

\section*{Acknowledgment}
This work was supported by MEXT Quantum Leap Flagship Program Grant No. JPMXS0120351339.

\end{document}